# Deep Convolutional Neural Network based Classification of Alzheimer's Disease using MRI Data

Ali Nawaz
Department of Software Engineering
University of Engineering and
Technology Taxila, Pakistan
nawazktk99@gmail.com

Syed Muhammad Anwar
Department of Software Engineering
University of Engineering and
Technology Taxila, Pakistan
s.anwar@uettaxila.edu.pk

Rehan Liaqat
Department of Electrical Engineering
University of Engineering and
Technology Taxila, Pakistan
rehan.liaqat@students.uettaxila.edu.pk

Javid Iqbal
Department of Software Engineering
University of Engineering and
Technology Taxila, Pakistan
javidiqbalkpk@gmail.com

Ulas Bagci
Center for Research in Computer
Vision, University of Central Florida,
Orlando, Florida, USA
bagci@ucf.edu

Muhammad Majid
Department of Computer Engineering,
University of Engineering and
Technology, Taxila, Pakistan
m.majid@uettaxila.edu.pk

*Abstract*—Alzheimer's disease (AD) is a progressive and incurable neurodegenerative disease which destroys brain cells and causes loss to patient's memory. Early detection can prevent the patient from further damage to the brain cells and hence avoid permanent memory loss. In the past few years, various automatic tools and techniques have been proposed for the diagnosis of AD. Several methods focus on fast, accurate, and early detection of the disease to minimize the loss to a patient's mental health. Although machine learning and deep learning techniques have significantly improved medical imaging systems for AD by providing diagnostic performance close to the human level. But the main problem faced during multi-class classification is the presence of highly correlated features in the brain structure. In this paper, we have proposed a smart and accurate way of diagnosing AD based on a two-dimensional deep convolutional neural network (2D-DCNN) using an imbalanced three-dimensional MRI dataset. Experimental results on Alzheimer's Disease Neuroimaging Initiative magnetic resonance imaging (MRI) dataset confirms that the proposed 2D-DCNN model is superior in terms of accuracy, efficiency, and robustness. The model classifies MRI into three categories: AD, mild cognitive impairment, and normal control; and has achieved 99.89% classification accuracy with imbalanced classes. The proposed model exhibits noticeable improvement in accuracy as compared to state-of-the-art methods.

## I. INTRODUCTION

Alzheimer's disease (AD) is a neurological brain disorder that affects the mental health of the people. The affected patients lose their ability of thinking, reading, and writing and could even forget the names of their family members in extreme cases. AD is the most common type of dementia that occurs in persons having signs of containing beta-amyloid in their brain cells, and as a result, such persons could need a full-time assistant to look after them. The pervasiveness of AD is calculated to be 5% in 65-year-old people and this percentage is shockingly higher (around 30%) for people of 85 years or older in developed countries [1]. The number of AD patients is expected to progress rapidly as life expectation upsurges globally. According to a recent study, AD is the 6[th] leading cause of death in the United States [1]. Almost 16 million Americans are offering unpaid caring of around 18.5 billion hours of worth $234 billion to 6 million American people suffering from AD [1]. The death rate due to heart attack in the last two decades has decreased by 9%, while the death ratio due to Alzheimer's has increased by 145% [1]. Among these only 16% of elderly people end up receiving proper care and routine check-ups. The cost of Alzheimer's and other dementia diseases in 2019 went up to 290 billion which is a major loss and in every 65 seconds, someone suffers from AD or dementia [1, 2]. There are three stages that lead up to AD –normal healthy control (NC), mild cognitive impairment (MCI), and Alzheimer's disease (AD). Accurate detection of AD is not possible until the patient's initial stage of dementia is converted to MCI.

Early recognition and apposite cure can stop the patient from reaching severe AD. One of the most common factors which drastically affects the human brain is shrinking of the brain cortex. The cortex area of the affected person shrivels whereas, normal shrinking happens mostly in the hippocampus area. This area provides the ability of thinking, memorizing of daily life activities, and a decrease in the hippocampus area leads to a shrink-up of brain cortex and enlargement of the ventricles. There are multiple ways to diagnose this disease which requires complete clinical information such as detailed history, physical and neurobiological exam, Neuropsychiatric Inventory-Questionnaire (NPI-Q), Functional Assessment Questionnaire (FAQ), Clinical Dementia Rating (CDR), Mini-Mental State Examination (MMSE), Global Deterioration Scale (GDS), and several other clinical evaluation parameters, developed for the diagnosis of AD by National Institute of Aging Alzheimer's Association [3]. Recent studies have found significant performance in the detection and classification of AD from multimodality data. Multimodality data contains Positron Emission Tomography (PET), Computed Tomography (CT), Magnetic Resonance Imaging (MRI), X-rays, and clinical records of the patient [4]. Although, MRI is still seen to be more effective in diagnosing AD than CT scans when using conventional machine learning, but several distinct problems are still associated with these techniques. Some of the problems are: (I) these techniques are time-consuming and require a lot of diagnosis experience (for labeling of the data) for an accurate diagnosis; especially in early stages. (II) pre-processing steps for manual feature extraction in these supervised methods could be error-prone and extracting low-level features from multiple imaging modalities could fail to get the best performance. However, a brain MRI represents a 3-D image volume containing information that could be beneficial for extracting features and training deeper convolutional neural networks.

To this end, we have proposed a two-dimensional deep convolution neural network (2D-DCNN) model which takes 3D MRI images as input for its training. The proposed model not only improves the classification accuracy but also addresses the class imbalance problem which is inherent in the multiclass classification problems. Our model can classify and detect three different stages of a subject i.e., NC, MCI, and AD. Furthermore, we have studied different neural network architectures like VGGNet and AlexNet for image classification; and have tuned our model yielding best results using convolution operation on medical imaging data.

The rest of this paper is organized as follows. In section 2, we provide a detailed literature study on multi-model and multi-class classification methods that have been presented in the past few years. Section 3 describes data selection and pre-processing steps of MRI data which is used as input to deep learning approaches. Section 4 provides an experimental setup, details of hyperparameters, model description, and performance evaluation of the 2D-CNN using 3D data. Conclusion and future work are discussed in section 5.

## II. LITERATURE REVIEW

AD and neurodegenerative diseases are progressive dementias. Dementia is a broader term for diseases caused by brain injury or diseases that adversely affect memory, thinking, and behavior. The disease causes permanent brain damage and ongoing effects on the human brain. According to the National Institute of Health (NIH) and Alzheimer's Association, 50-80% of dementia patients convert to AD [1, 5]. There is no proper treatment for AD and most of the subjects are diagnosed after 65-70 years. Although, early detection and diagnosis could prevent the impact of disease progression. In the past few years, significant work is done for the diagnosis and automatic classification of AD. In [6], dimension reduction methods and their variations were used with whole-brain MRI data for the binary and multi-class classification and detection of AD. A hybrid approach was proposed for the classification of AD in its prodromal stage and its accuracy was improved using a hybrid feature vector constructed from MRI and clinical data [2]. The features were extracted from the whole magnetic resonance images as well as from the segmented regions demonstrating grey matter (GM), white matter (WM), and cerebrospinal fluid (CSF). The grey level co-occurrence matrix (GLCM) had the best results among other methods. One step DCNN model was implemented in [7] for detection and automatic classification of AD on the Open Access Series of Imaging Studies (OASIS) dataset. In [8], AD was detected and classified by developing an ensemble of DCNN using the OASIS dataset which identified current and early stages of Alzheimer's disease patient. Reference [9] succeeded to achieve a high level of accuracy using 1.5T T1-weighted images from ADNI as well as from the recruited patient from their institute by developing deep learning algorithm which diagnosed patients with AD and patients where MCI was converted to AD or chronic MCI based on single cross-sectional MRI. In [10], a novel method was presented to find an objective measure for eliminating MCI subjects from those who were a greater risk to convert from MCI to AD using several modalities like; MRI, neuropsychological and clinical data, and APOe4 genetic classification. The model used the transfer leaning method for capturing genetic features using pre-trained ImageNet and ResNet 152 models.

Machine learning and deep learning approaches are one of the basic needs of computer vision and data science researchers now a days. Deep learning approaches learn unique demonstrations from raw data due to their hierarchal and layer-wise structure. A convolutional neural network (CNN) is inspired from the human brain which learns features from data through a hierarchical method going from simple to complex representation. In [11], deep learning techniques were applied for disease diagnosis using medical images and significant results were obtained for recognition, classification, and segmentation. Replacement of conventional methods by deep learning approaches improved the classification results from 84.4% to 96% for an unbiased dataset [12]. In [13], a multi-class graph CNN (G-CNN) was used for the classification of AD data which enhanced the classification performance of CNN. Training and validation of the network were done using two methods: (a) structural connectivity graphs which are based on diffusion tensor imaging (DTI) (b) G-CNN classifier using receiver operating characteristics (ROC) analysis which performs better than traditional support vector machine (SVM) classifier. Particle swarm optimization (PSO) technique was used to extract features to analyze changes in the brain structure which is further related to the clinical progression of AD by applying a multi-class classifier to achieve the best performance of recognition and classification [14].

A novel method was presented in [15] to figure out specific texture area in images by applying the following steps: (a) extract texture features from 3D MRI images and divide them into 2D slices in different axes and mark the region of interest (ROI), (b) use fisher ranking, elastic net and SVM to eliminate recursive features (c) eliminated features were further analyzed and classified using multiple classifiers i.e., SVM, K-nearest neighbors, and random forest. The experiment was performed on 812 subjects from the ADNI data and achieved improved accuracy as compared to the state-of-the-art methods. A similar approach used the idea of texture-based segmentation of different sections of MRI images of the OASIS dataset and used k-means method to separate CSF, GM, and WM [16]. These feature vectors from each segmented region were combined to form a single feature vector containing texture- and shape-based features which were further used by SVM, KNN, and random forest to classify AD. In [17], detection and classification of AD were performed by deep learning using MRI data to reduce the cost of diagnosis and caring for AD patients. A 3D-CNN with 39 layers was used, which automatically learned features and used a residual connection to prevent feature loss and achieved a high level of accuracy using ResNet with 52 layers.

Currently, deep learning techniques are being used in most studies since they drastically achieve improved results. Class imbalance is a common problem in the biomedical imaging field which hinders accurate diagnosis and classification of the disease. Recent studies in the literature have successfully applied CNN on 2D and 3D data to achieve promising results for binary and multi-class classification problems. Most of the deep learning methods are now replacing fully connected layers in dense networks by convolution layers. A deep CNN model was trained on a small imbalanced class dataset to classify AD and its different stages [18]. In [19], the class imbalance was dealt for segmentation using a focal loss function which reshaped the cross-entropy loss function with a modulating exponent to down-weight errors assigned to well-classified instances. This focal loss function prevented a negative gradient for a large quantity from dominating the learning process.

Existing models for AD classification presented in the literature are complex as they generally employ 3D MRI data and use complete information of highly correlated features which increases the computational cost and limits the effectiveness of multi-class classification. Class imbalance is another problem generally faced during diagnosis studies which are handled generally by augmentation. To overcome the aforementioned issues, in this paper, we have proposed a smart and accurate way of diagnosing different stages of AD based on a 2D-DCNN model using imbalanced 3D MRI data. The contributions of this work are threefold: (i) Instead of using 3D MRI data, we first convert 3D images into 2D slices, and hence simplify the structure of DCNN model. (ii) Instead of using all 2D slices, the proposed model uses selected 2-D slices with promising features as its input to reduce computational burden (iii) we effectively handle the problem of imbalanced data without using any augmentation. Simulation studies are carried out on a subset of ADNI MRI dataset to evaluate the accuracy, efficiency, and robustness of the method. Furthermore, results are compared with recent studies available in the literature for validation of the proposed model.

## III. MATERIALS AND METHODOLOGY

### A. Data Selection

In this study, we have used a subset of the Alzheimer's Disease Neuroimaging Initiative (ADNI) (http://www.loni.ucla.edu/ADNI) data which can be accessed and obtained from their database [5]. The dataset contains multiple modality data e.g. PET, functional MRI, MRI, genetic data, and clinical information for thousands of patients. Structural MRI volumes were acquired from the subjects on 1.5T and 3T scanners with a multi-time period with a gap of one to three years. The ADNI project was started in 2003 with the collaboration of the National Institute of Aging, Bioengineering (NIBI), Biomedical Imaging, Private Pharmaceutical Organization, Food and Drug Administration (FDA), and some non-profit companies for 5-year partnership with the worth of $60 million. The aim of this platform is to validate whether different modality data combination could measure the progress of AD and its different stages.

We found that for each patient, there are normally 2-6 images corresponding to them, with different ages (but super close and the rest of the information are the same). Hence, we could pick one image for each person to avoid overfitting and redundant features. Therefore, the earliest image corresponding to an individual patient was selected. In our study, we have used 3D brain MR images of 160 subjects (52 NC, 62 MCI, and 45 AD) to train our 2D-CNN model on a total number of 67413 to diagnose AD. The unbalanced dataset includes 20972 images for AD class, 26192 images for MCI, and 18513 for CN class. Other demographic information from ADNI-1 and the subset of data we have used in this study are given in Table. I.

### B. MRI Preprocessing

MRI acquisition contains noisy data which must be accounted for before computational analysis of the data. This raw data needs noise removal, intensity normalization, contrast adjustment, and removal of background area which are not necessary for processing. The dataset has followed initial pre-processing steps during its acquisition in terms of Gradwarp, B1 non-uniformity, and N3 bias field correction. Three-dimensional data which is used in this work consists of sagittal, transversal, and coronal views of the subject. Every MPRAGE (Magnetization Prepared Rapid Acquisition Gradient Echo) scan in the ADNI database connected to corresponding MRI image files goes through standard acquisition steps. MRI images are provided in NII format (Neuroimaging Informatics Technology Initiative - Nifti) for all the three stages with different resolutions. Therefore, some pre-processing steps have been applied in order to feed the data to our model. First of all, spatial MRI images are normalized to validate the position of the image. Next, intensity normalization, noise removal, bias correction, contrast adjustment, and rescaling was performed using the MANGO tool kit [20] as shown in Fig. 1. Every 3D MRI image contains $256 \times 256 \times 166$ slices per volume which cannot be fed to a 2D CNN model. Therefore, we have rescaled each 3D MRI volume and have converted it into 2D slices each of size $300 \times 300$ with a single channel for each plane (axial, coronal, sagittal). Each patient contains around $690\pm$ 2D slices which can be further fed to train the 2D-CNN model. The pre-processed slices of 3D images are shown in Fig. 1 during different stages.

Table I: Demographic information of subjects from ADNI dataset used in this study.

| Classes | Total Volumes | Age | Sex | No of Volumes | No of used slices |
|---|---|---|---|---|---|
| CN | 574 | (55-98)± | 29(M), 24(F) | 53 | 20,972 |
| MCI | 812 | (45-105)± | 43(M), 19(F) | 62 | 26,192 |
| AD | 340 | (50-102)± | 22(M), 23(F) | 45 | 18,513 |
| Total | 1726 | (45-105)± | 94(M), 66(F) | 160 | 65,677 |

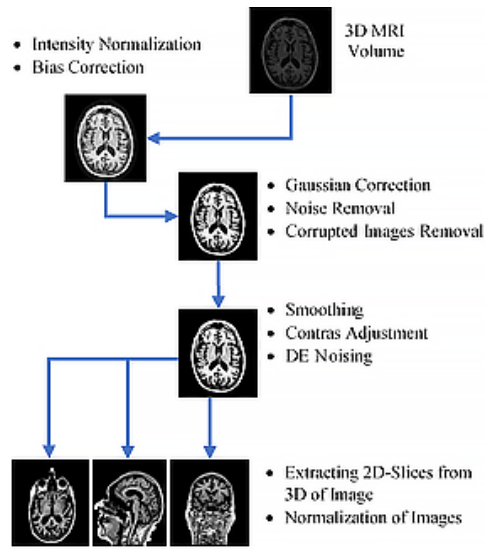

Fig. 1: An overview of the pre-processing steps.

## C. Deep Networks Architecture

The CNN architecture, we used for this study is inspired from the human visual cortex. The human eye receives the input stream of information in its receptive field which is similar to the convolution operation that convolves the input image and makes the feature map by operating on its input receptive field. Convolution operation includes multiple layers which contain ReLU activation functions, max-pooling layers, and fully connected layers. Every input is passed through these operations to reach the final output in the form of binary or multi-class classifier. The convolution operation is connected through a bunch of neurons, shared hyper-parameters, local connectivity, and shift-invariance which makes the network more powerful. Multiple CNN architectures including Alex Net, VGG-Net, GoogleNet, ResNet have been deployed in the literature to perform classification tasks on large image datasets. These models had been successfully used in various applications like image classification, recognition, image captioning, pose identification [21]. The subset ADNI dataset i.e. ADNI-1 used in this work for simulation is also evaluated using Alex Net, VGG-16, and the proposed Deep ConvNet. The details of these models are given below for completeness.

### 1. Alex Net

AlexNet introduced for the ImageNet competition consisted of 5 convolutional layers which are followed by overlapping max-pooling layers. Further, three fully connected layers are present at the end with an input size of $224 \times 224 \times 3$ [22]. This complex deep learning network was trained on 1.2 million high-resolution images from 1000 different classes. Different sizes of kernel and filters extract complex information from fine-grained large-scale images, while the overlapping max pooling and dropout layers reduce the dimensionality [22]. In this study, we applied the original Alex Net architecture to classify multiple stages of AD using the ADNI-1 dataset.

### 2. VGG-16

VGG-16 is a deep convolutional neural network which won ILSVR-2013 and substantially outperformed GoogLeNet [23]. The model achieved an accuracy of 92.7% on 14 million images belonging to 1000 classes. The input to the networks included fixed size ($224 \times 224 \times 3$) RGB images which were further passed through a sequential convolutional layer. Very small receptive filters with a size of $3 \times 3$ and $1 \times 1$ were applied which significantly improved results as compared to prior studies by pushing depth of the weight layers from 16-19. These convolutional layers followed by three fully connected layers with 4096 units and the output layer used SoftMax activation with 1000 units for classification [23]. VGG-16 is used as a baseline model in this study and results are improved on the ADNI-1 dataset to classify multiple stages of AD.

### 3. Proposed Deep ConvNet Model

Convolutional layers are basic building blocks for any deep CNN reaching the best performance with applying complex activation functions. The proposed deep CNN (deep ConvNet) model automatically extracts features from whole-brain MRI scans with a deep convolutional neural network to diagnose Alzheimer's disease. The proposed model is shown in Fig. 2, where pre-processed data with 2D slices are fed as input. Each subject contains multiple (590 - 690) slices per volume. We applied 6 repeated blocks of 2D convolutional layers with the same padding size and $300 \times 300 \times 1$ image size. Kernel size for all the convolutional layers was set to be the same, however, 4 to 128 size filters were used to extract complex and multi-scale information which makes a feature map to pass these features to the next layers for extracting further complex features. CNN architecture works on filters i.e., first few filters learn edges in the image, next filters learn complex shapes of the image and subsequent filters extract colors and complex structures. CNN architecture tracks the computational power of the input layer by incorporating a reduction of dimensionality to input shape by max pooling. The size of the filter in each convolutional layer is also varied. The input image is converted into a feature map that passes through a concatenation process called fully connected layers. We have used five fully connected layers with 512, 256, 128, 64, 32 and 16 neurons in FC1, FC2, FC3, FC4, FC5 and FC6, respectively. The classification layer contains 3 neurons according to the number of output classes.

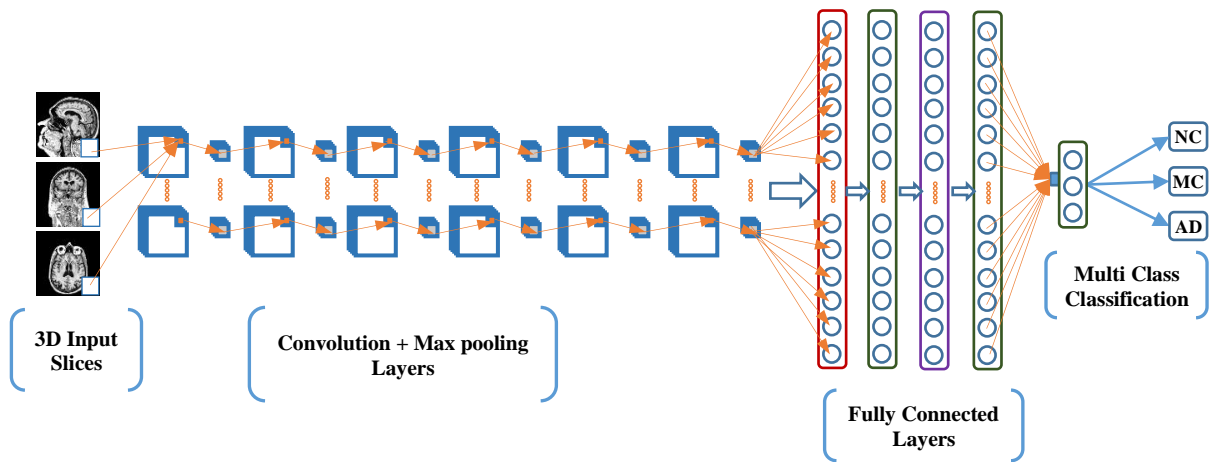

Fig. 2: Our proposed convolutional neural network (CNN) model.

## IV. EXPERIMENTAL RESULTS AND DISCUSSION

In this paper, we have used 3D structural MRI scans of 160 patients (52 NC, 62 MCI, and 45 AD) to train our 2D-CNN model. The unbalanced (a total of 67413) 2D images are used as a dataset which includes 20972 images for AD class, 26192 images for MCI, and 18513 for NC class. Networks are trained from scratch on data for 70 epochs with a batch size of 100. Experiments are performed using 60% data for training, 20% for testing, and 20% for the validation set. The implementation is performed using Keras deep learning library with Anaconda Jupiter Notebook. Every convolutional layer uses zero-padding with a 2x2 stride and a ReLU activation function for better convergence. The input image with $300 \times 300$ size is followed by 5 fully connected layers with 1024, 512, 256, 128, 64 and 32 neurons in FC1, FC2, FC3, FC4, FC5 and FC6, respectively. MaxPooling with $2\times 2$ stride is used for reducing dimensionality and extracting spatial features. Another reason to use max-pooling is that it gives the highest possible average values during pooling of the convolution feature map which boosts the accuracy rate. To reduce the time for the selection of hyperparameters, we have used standard CNN hyper-parameters with fine-tuning of the model. The network is optimized by using RMSProp optimizer and accuracy as loss function with a learning rate of 0.0001 which gives the best classification results with a smaller batch size to save memory consumption. We have used categorical cross-entropy loss function due to multi-label input data. The final SoftMax layer takes feature representation $f_i$ from adjacent fully connected layers and interpret the output class.

A probability score $p_i$ is assigned to any $i_{th}$ output class where each class defines each unique stage of the disease. For generalization, if we assume $n$ different stages of the disease, we can write $p_i$ as equation (1). In our case, $n =3$ as we have used three stages (AD, MCI, and NC). Similarly, categorical entropy loss for network '$L$' can be found by using ground truth '$t_i$' for MRI and $p_i$ as shown in equation (2), while the difference between probability score per class '$p_i$' and ground truth '$t_i$' was represented by equation (3).

$$p_i = \frac{exp^{(fi)}}{\sum_{j=0}^{n} exp^{(fi)}}, i = 1,2,3,\dots\dots,n \quad (1)$$

$$L = - \sum_{i=0}^{n} t_i . \log(p_{i,}) \quad (2)$$

$$\frac{\partial L}{\partial f_i} = p_i - t_i \quad (3)$$

The optimal learning rate is selected to calculate gradient per epoch during backpropagation operation. There are multiple subsets to select hyper-parameters for any neural network. The model has better features detection to classify larger datasets. Our target is to classify AD stages on imbalanced data, and we have successfully achieved it with the accuracy of 98.97% and 99.31% on Alex Net and VGG-16 respectively. Whereas, the proposed model, with a small number of layers than VGG, benefits the training time and computational burden, performs best in terms of accuracy of 99.89% which is superior to other state-of-the-art techniques presented in literature. However, most of the studies used a combination of features from different modalities, while the proposed method is more suitable for time-saving with single modality and low-level pre-processing. The results presented in the form of confusion matrix, the loss values, and accuracy are shown in Figures 3, 4, and 5 respectively. Table II shows the comparison results of previous studies with the proposed framework. The confusion matrix and ROC curve on training, testing, and validation data demonstrate the effectiveness of this model. This paper serves as a further stimulus for utilizing different deep learning-based models for the analysis of 3D MRI images in the detection of AD and its earlier stages.

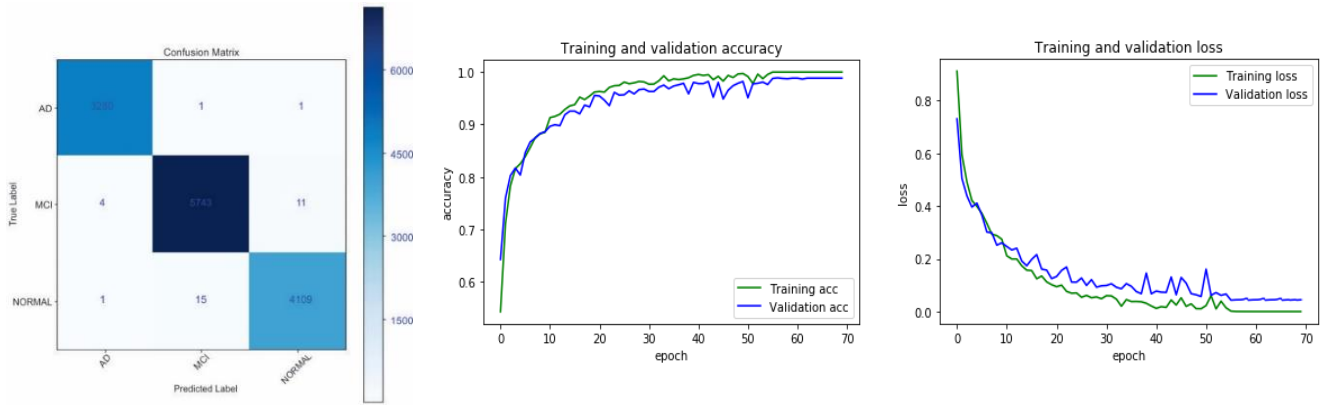

Fig.3 (Right Image) Confusion matrix for testing data with true and predicted labels, (Middle Image) Accuracy for training and testing data on 70 epochs (Right Side) Loss for training and validation data.

TABLE II: CLASSIFICATION PERFORMANCE OF PROPOSED MODELS WITH RESULTS OF THE STATE-OF-THE-ART TECHNIQUES.

| Models | Architecture/Methodology | Modality | Subjects | 3-way Classification | Accuracy % |
|---|---|---|---|---|---|
| [2] | GLCM | sMRI + Clinical | 287 | 3 way | 79.9% |
| [6] | Voxel/Hippo | MRI | 509 | 3 way | 95% |
| [7] | Deep CNN | MRI | 416 | 4 way (AD, EMCI, LMCI, NC) | 73.75% |
| [16] | K-Means/SVM/RF | MRI | 128 | 2/3 way | 92.7% |
| [24] | CNN | MRI | 416 | 3 way | 96.25% |
| [12] | CNN | MRI | Subset | 3 way | 96% |
| [25] | DCNN | MRI | 149 | 4 way | 98.88% |
| **Baseline** | VGG-16 | MRI | 160 | 3 way | **98.97%** |
| **Baseline** | Alex Net | MRI | 160 | 3 way | **99.31%** |
| **Proposed Model** | Deep ConvNet | MRI | 160 | 3 way | **99.89%** |

## V. CONCLUSION

Most studies classify multiclass data by incorporating multimodality features and pre-trained feature learning methods. In this study, we presented a deep ConvNet based model for classifying MRI images to detect and classify AD, MCI, and NC. The imbalance data with 160 volumes (45 AD, 62 MCI, and 52 NC) was provided by ADNI: whole and pre-processed MRI scans were passed to the proposed DCNN. The proposed network was compared with baseline models including Alex Net and VGG-16. We achieved an accuracy of 99.89% in the 3-class classification of Alzheimer's disease, which is superior to other frameworks presented in the literature. The proposed model gives an accurate prediction of the classes and proves that incorporating deep ConvNet models helps learn distinctive features from neuroimaging data. The proposed method has a high level of implications in neuro- as well as other modalities in medical imaging. This work can be further improved by combining metadata and clinical details for an individual subject to cross-check the results obtained through clinical and MRI data.